%% file: arxiv.tex
\pdfoutput=1

\documentclass[12pt,a4paper]{article}

\usepackage{ifthen} 
\newboolean{pdflatex}
\setboolean{pdflatex}{true} 

\newboolean{articletitles}
\setboolean{articletitles}{true} 

\newboolean{uprightparticles}
\setboolean{uprightparticles}{false} 

\newboolean{inbibliography}
\setboolean{inbibliography}{false} 

\def\paperauthors{LHCb collaboration} 
\def\paperasciititle{Search for baryon-number-violating Xi_b^0 oscillations} 
\def\papertitle{Search for baryon-number-violating \Xibz oscillations} 
\def\paperkeywords{{High Energy Physics}, {LHCb}} 
\def\papercopyright{CERN on behalf of the LHCb collaboration}
\def\paperlicence{CC-BY-4.0}
\def\paperlicenceurl{https://creativecommons.org/licenses/by/4.0/}

\input{preamble}
\usepackage{longtable} 

\input{xib-symbols-def}
\def\bnvsubplotwidth{0.8\textwidth}
\def\bnvplotwidth{0.8\textwidth}

\begin{document}

\renewcommand{\thefootnote}{\fnsymbol{footnote}}
\setcounter{footnote}{1}

\input{title-LHCb-PAPER}

\renewcommand{\thefootnote}{\arabic{footnote}}
\setcounter{footnote}{0}

\pagestyle{plain} 
\setcounter{page}{1}
\pagenumbering{arabic}


\input{bnv-paper-body.tex}

\input{acknowledgements}

\addcontentsline{toc}{section}{References}
\setboolean{inbibliography}{true}
\bibliographystyle{LHCb}
\bibliography{main,LHCb-PAPER,LHCb-CONF,LHCb-DP,LHCb-TDR,bnv-refs}

\newpage
 
\input{LHCb_Authorship_flat_14-Jun-2017.tex}

\end{document}

%% file: preamble.tex
\usepackage[top=1in, bottom=1.25in, left=1in, right=1in]{geometry}

\usepackage{microtype}
\usepackage{lineno}  
\usepackage{xspace} 
\usepackage{caption} 

\usepackage{graphicx}  
\usepackage{color}
\usepackage{colortbl}
\graphicspath{{./figs/}} 

\usepackage{amsmath} 
\usepackage{amssymb}
\usepackage{amsfonts}
\usepackage{upgreek} 

\newcommand*\patchAmsMathEnvironmentForLineno[1]{%
\expandafter\let\csname old#1\expandafter\endcsname\csname #1\endcsname
\expandafter\let\csname oldend#1\expandafter\endcsname\csname
end#1\endcsname
 \renewenvironment{#1}%
   {\linenomath\csname old#1\endcsname}%
   {\csname oldend#1\endcsname\endlinenomath}%
}
\newcommand*\patchBothAmsMathEnvironmentsForLineno[1]{%
  \patchAmsMathEnvironmentForLineno{#1}%
  \patchAmsMathEnvironmentForLineno{#1*}%
}
\AtBeginDocument{%
\patchBothAmsMathEnvironmentsForLineno{equation}%
\patchBothAmsMathEnvironmentsForLineno{align}%
\patchBothAmsMathEnvironmentsForLineno{flalign}%
\patchBothAmsMathEnvironmentsForLineno{alignat}%
\patchBothAmsMathEnvironmentsForLineno{gather}%
\patchBothAmsMathEnvironmentsForLineno{multline}%
\patchBothAmsMathEnvironmentsForLineno{eqnarray}%
}

\usepackage{hyperxmp}

\usepackage[pdftex,
            pdfauthor={\paperauthors},
            pdftitle={\paperasciititle},
            pdfkeywords={\paperkeywords},
            pdfcopyright={Copyright (C) \papercopyright},
            pdflicenseurl={\paperlicenceurl}]{hyperref}

\usepackage[all]{hypcap} 

\input{lhcb-symbols-def} 

\usepackage{cite} 
\usepackage{mciteplus}

%% file: lhcb-symbols-def.tex
\usepackage{xspace} 
\usepackage{upgreek}

\def\lhcb {\mbox{LHCb}\xspace}

\def\MagUp {\mbox{\em Mag\kern -0.05em Up}\xspace}

\ifthenelse{\boolean{uprightparticles}}%
{

 \def\Ppi         {\ensuremath{\uppi}\xspace}

 \def\PDelta      {\ensuremath{\Delta}\xspace}                 
 \def\PXi      {\ensuremath{\Xi}\xspace}                 
 \def\PLambda      {\ensuremath{\Lambda}\xspace}                 
 \def\PSigma      {\ensuremath{\Sigma}\xspace}                 
 \def\POmega      {\ensuremath{\Omega}\xspace}                 
 \def\PUpsilon      {\ensuremath{\Upsilon}\xspace}                 
 
 \def\PB      {\ensuremath{\mathrm{B}}\xspace}                 
                  
 \def\PD      {\ensuremath{\mathrm{D}}\xspace}

 \def\PK      {\ensuremath{\mathrm{K}}\xspace}

 \def\Pb      {\ensuremath{\mathrm{b}}\xspace}                 
 \def\Pc      {\ensuremath{\mathrm{c}}\xspace}

 \def\Pi      {\ensuremath{\mathrm{i}}\xspace}

 \def\Pp      {\ensuremath{\mathrm{p}}\xspace}

 \def\Ps      {\ensuremath{\mathrm{s}}\xspace}

}
{

 \def\Ppi         {\ensuremath{\pi}\xspace}

 \mathchardef\PDelta="7101
 \mathchardef\PXi="7104
 \mathchardef\PLambda="7103
 \mathchardef\PSigma="7106
 \mathchardef\POmega="710A
 \mathchardef\PUpsilon="7107
                  
 \def\PB      {\ensuremath{B}\xspace}                 
                  
 \def\PD      {\ensuremath{D}\xspace}

 \def\PK      {\ensuremath{K}\xspace}

 \def\Pb      {\ensuremath{b}\xspace}                 
 \def\Pc      {\ensuremath{c}\xspace}

 \def\Pi      {\ensuremath{i}\xspace}

 \def\Pp      {\ensuremath{p}\xspace}

 \def\Ps      {\ensuremath{s}\xspace}

}

\makeatletter
\ifcase \@ptsize \relax
  \newcommand{\miniscule}{\@setfontsize\miniscule{4}{5}}
\or
  \newcommand{\miniscule}{\@setfontsize\miniscule{5}{6}}
\or
  \newcommand{\miniscule}{\@setfontsize\miniscule{5}{6}}
\fi
\makeatother

\DeclareRobustCommand{\optbar}[1]{\shortstack{{\miniscule (\rule[.5ex]{1.25em}{.18mm})}
  \\ [-.7ex] $#1$}}

\def\squark    {{\ensuremath{\Ps}}\xspace}

\def\cquark    {{\ensuremath{\Pc}}\xspace}

\def\bquark    {{\ensuremath{\Pb}}\xspace}

\def\pion   {{\ensuremath{\Ppi}}\xspace}

\def\pip    {{\ensuremath{\pion^+}}\xspace}
\def\pim    {{\ensuremath{\pion^-}}\xspace}

\def\kaon    {{\ensuremath{\PK}}\xspace}
  \def\Kbar    {{\kern 0.2em\overline{\kern -0.2em \PK}{}}\xspace}

\def\KorKbar    {\kern 0.18em\optbar{\kern -0.18em K}{}\xspace}
\def\Kz      {{\ensuremath{\kaon^0}}\xspace}

\def\Kp      {{\ensuremath{\kaon^+}}\xspace}
\def\Km      {{\ensuremath{\kaon^-}}\xspace}

  \def\Dbar    {{\kern 0.2em\overline{\kern -0.2em \PD}{}}\xspace}
\def\D       {{\ensuremath{\PD}}\xspace}

\def\DorDbar    {\kern 0.18em\optbar{\kern -0.18em D}{}\xspace}
\def\Dz      {{\ensuremath{\D^0}}\xspace}

\def\B       {{\ensuremath{\PB}}\xspace}
\def\Bbar    {{\ensuremath{\kern 0.18em\overline{\kern -0.18em \PB}{}}}\xspace}

\def\BorBbar    {\kern 0.18em\optbar{\kern -0.18em B}{}\xspace}
\def\Bz      {{\ensuremath{\B^0}}\xspace}

\def\Bs      {{\ensuremath{\B^0_\squark}}\xspace}

  \def\Y#1S{\ensuremath{\PUpsilon{(#1S)}}\xspace}

\def\proton      {{\ensuremath{\Pp}}\xspace}
\def\antiproton  {{\ensuremath{\overline \proton}}\xspace}

\def\Xires       {{\ensuremath{\PXi}}\xspace}
\def\Xiresbar    {{\ensuremath{\overline \Xires}}\xspace}

\def\Lbar        {{\ensuremath{\kern 0.1em\overline{\kern -0.1em\PLambda}}}\xspace}
\def\LorLbar    {\kern 0.18em\optbar{\kern -0.18em \PLambda}{}\xspace}

\def\Xibz    {{\ensuremath{\Xires^0_\bquark}}\xspace}

\def\Xibbarz {{\ensuremath{\Xiresbar{}_\bquark^0}}\xspace}

\def\Xicp    {{\ensuremath{\Xires^+_\cquark}}\xspace}

\def\Xicbarm {{\ensuremath{\Xiresbar{}_\cquark^-}}\xspace}

\def\to                 {\ensuremath{\rightarrow}\xspace}

\def\CP                {{\ensuremath{C\!P}}\xspace}

\def\AT#1     {\ensuremath{A_{\mathrm{T}}^{#1}}\xspace}           

\def\C#1      {\ensuremath{\mathcal{C}_{#1}}\xspace}                       
\def\Cp#1     {\ensuremath{\mathcal{C}_{#1}^{'}}\xspace}                    
\def\Ceff#1   {\ensuremath{\mathcal{C}_{#1}^{\mathrm{(eff)}}}\xspace}        
\def\Cpeff#1  {\ensuremath{\mathcal{C}_{#1}^{'\mathrm{(eff)}}}\xspace}       
\def\Ope#1    {\ensuremath{\mathcal{O}_{#1}}\xspace}                       
\def\Opep#1   {\ensuremath{\mathcal{O}_{#1}^{'}}\xspace}                    

\newcommand{\tev}{\ifthenelse{\boolean{inbibliography}}{\ensuremath{~T\kern -0.05em eV}}{\ensuremath{\mathrm{\,Te\kern -0.1em V}}}\xspace}
\newcommand{\gev}{\ensuremath{\mathrm{\,Ge\kern -0.1em V}}\xspace}
\newcommand{\mev}{\ensuremath{\mathrm{\,Me\kern -0.1em V}}\xspace}
\newcommand{\kev}{\ensuremath{\mathrm{\,ke\kern -0.1em V}}\xspace}
\newcommand{\ev}{\ensuremath{\mathrm{\,e\kern -0.1em V}}\xspace}
\newcommand{\gevc}{\ensuremath{{\mathrm{\,Ge\kern -0.1em V\!/}c}}\xspace}
\newcommand{\mevc}{\ensuremath{{\mathrm{\,Me\kern -0.1em V\!/}c}}\xspace}
\newcommand{\gevcc}{\ensuremath{{\mathrm{\,Ge\kern -0.1em V\!/}c^2}}\xspace}
\newcommand{\gevgevcccc}{\ensuremath{{\mathrm{\,Ge\kern -0.1em V^2\!/}c^4}}\xspace}
\newcommand{\mevcc}{\ensuremath{{\mathrm{\,Me\kern -0.1em V\!/}c^2}}\xspace}

\def\invfb   {\ensuremath{\mbox{\,fb}^{-1}}\xspace}

\def\ps   {\ensuremath{{\mathrm{ \,ps}}}\xspace}

\def\gsim{{~\raise.15em\hbox{$>$}\kern-.85em
          \lower.35em\hbox{$\sim$}~}\xspace}
\def\lsim{{~\raise.15em\hbox{$<$}\kern-.85em
          \lower.35em\hbox{$\sim$}~}\xspace}

\def\evtgen     {\mbox{\textsc{EvtGen}}\xspace}

\def\geant      {\mbox{\textsc{Geant4}}\xspace}

\def\photos     {\mbox{\textsc{Photos}}\xspace}

\def\pythia     {\mbox{\textsc{Pythia}}\xspace}

\def\tell1  {TELL1\xspace}
\def\ukl1   {UKL1\xspace}

\newcommand{\eg}{\mbox{\itshape e.g.}\xspace}

%% file: xib-symbols-def.tex
\def\XibxPrime  {{\ensuremath{\Xires^{\prime}_\bquark}}\xspace}
\def\XibxStar  {{\ensuremath{\Xires^{*}_\bquark}}\xspace}
\def\XibPrimeMinus  {{\ensuremath{\Xires^{\prime -}_\bquark}}\xspace}

\def\XibStarMinus  {{\ensuremath{\Xires^{*-}_\bquark}}\xspace}

\def\XibPrimeOrStarMinus  {{\ensuremath{\Xires^{\prime,* \, -}_\bquark}}\xspace}

\newcommand{\kevcc}{\ensuremath{{\mathrm{\,ke\kern -0.1em V\!/}c^2}}\xspace}

%% file: title-LHCb-PAPER.tex
\begin{titlepage}
\pagenumbering{roman}

\vspace*{-1.5cm}
\centerline{\large EUROPEAN ORGANIZATION FOR NUCLEAR RESEARCH (CERN)}
\vspace*{1.5cm}
\noindent
\begin{tabular*}{\linewidth}{lc@{\extracolsep{\fill}}r@{\extracolsep{0pt}}}
\ifthenelse{\boolean{pdflatex}}
{\vspace*{-2.7cm}\mbox{\!\!\!\includegraphics[width=.14\textwidth]{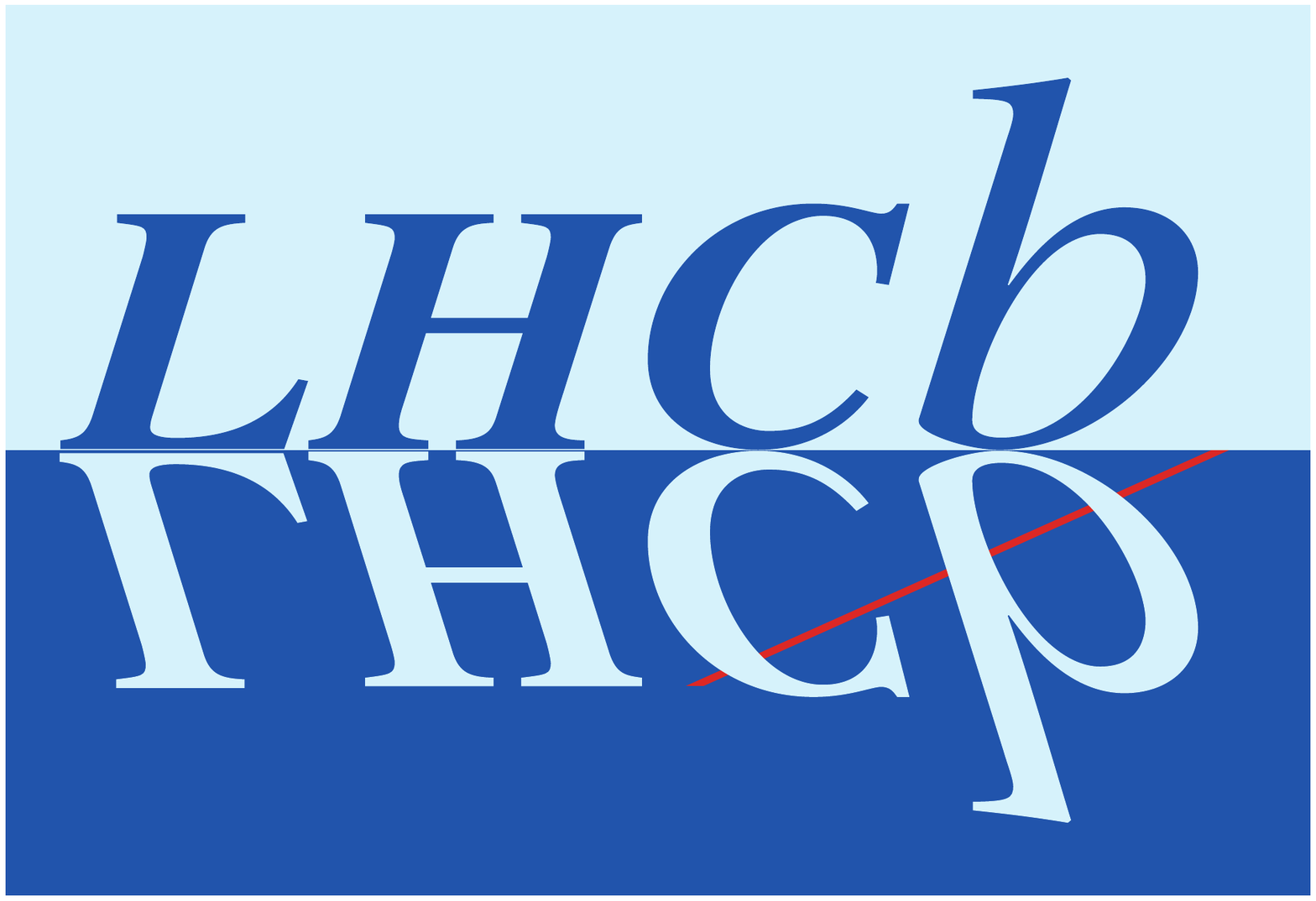}} & &}%
{\vspace*{-1.2cm}\mbox{\!\!\!\includegraphics[width=.12\textwidth]{lhcb-logo.eps}} & &}%
\\
 & & CERN-EP-2017-189 \\  
 & & LHCb-PAPER-2017-023 \\  
 & & November 4, 2017 \\ 
 & & \\
\end{tabular*}

\vspace*{4.0cm}

{\normalfont\bfseries\boldmath\huge
\begin{center}
  \papertitle 
\end{center}
}

\vspace*{2.0cm}

\begin{center}
\paperauthors\footnote{Authors are listed at the end of this Letter.}
\end{center}

\vspace{\fill}

\begin{abstract}
  \noindent
\input{abstract.tex}

\end{abstract}

\vspace*{2.0cm}

\begin{center}
  Published as \href{https://dx.doi.org/10.1103/PhysRevLett.119.181807}{Phys.~Rev.~Lett. \textbf{119}, 181807 (2017)}
\end{center}

\vspace{\fill}

{\footnotesize 
\centerline{\copyright~\papercopyright, licence \href{\paperlicenceurl}{\paperlicence}.}}
\vspace*{2mm}

\end{titlepage}


\newpage
\setcounter{page}{2}
\mbox{~}

\cleardoublepage

%% file: abstract.tex
 A search for baryon-number-violating \Xibz oscillations
 is performed
 with a sample of $\proton \proton$ collision data
 recorded by the LHCb experiment,
 corresponding to an integrated luminosity of 3\invfb.
  The baryon number at
  the moment of production is identified by requiring that the \Xibz
  come from the decay of a resonance ${\XibStarMinus \to \Xibz \pim}$
  or ${\XibPrimeMinus \to \Xibz \pim}$,
  and the baryon number at the moment of decay is identified from
  the final state using the decays $\Xibz \to \Xicp \pim, ~ \Xicp \to p \Km \pip$.
  No evidence of baryon number violation is found, and an upper limit
at the 95\% confidence level
  is set on the oscillation rate of $\omega < 0.08 \, \ps^{-1}$,
  where $\omega$ is the associated angular frequency.

%% file: bnv-paper-body.tex
Three conditions are necessary for the formation of a
matter-dominated universe:
$C$ and \CP violation,
baryon number violation~(BNV),
and the absence of thermal equilibrium~\cite{Sakharov:1967dj}.
The existence of $C$ and \CP violation has been established
experimentally for decades~\cite{Christenson:1964fg,Aubert:2001nu,Abe:2001xe}, although the amount of \CP violation
present in the Standard Model~(SM) is known to be insufficient
to generate the matter-antimatter asymmetry observed in the universe today~\cite{Gavela:1993ts,Gavela:1994dt},
prompting numerous searches for sources of \CP violation beyond
the SM. By contrast,
despite baryon number conservation being an accidental low-temperature symmetry of the SM,
BNV has never been
observed experimentally, and stringent lower limits have been placed
on the mean lifetimes of protons and of bound neutrons~\cite{PDG2017}.
These limits impose constraints on generic models of physics
beyond the SM.
In particular, in supersymmetric extensions of the SM,
a mechanism such as $R$-parity conservation is required
to naturally suppress baryon number violation~\cite{Salam:1974xa,Fayet:1974pd,Weinberg:1981wj,Farrar:1982te}.
An alternative is that the new physics has
nongeneric
flavour interactions, such that only certain BNV processes are
allowed and the experimental constraints are respected.
One possibility would be for
new BNV couplings to be entirely
flavour diagonal~\cite{Smith:2011rp,Durieux:2012gj}, such as
a six-fermion operator that
couples two fermions from each generation.
This would couple two from each of
$\{ u, d, e, \nu_e \}$, 
$\{ c, s, \mu, \nu_\mu \}$, and
$\{ t, b, \tau, \nu_\tau \}$, with
duplication allowed within a generation, \eg a
$usb\overline{usb}$
vertex would be permitted.
Such an operator could arise in models with leptoquarks or $R$-parity-violating
supersymmetric extensions of the SM~\cite{Durieux:2012gj,Aitken:2017wie}.
The six-fermion operator could allow BNV while being consistent with the experimental
limit on the proton lifetime, since the proton initial state contains
only first-generation fermions and, therefore, its coupling to the
operator would require two flavour-changing neutral processes and
would be heavily suppressed~\cite{Durieux:2012gj}.

Most experimental processes involving such an operator are difficult
to observe, since they include multiple third-generation fermions.
For example, the signatures proposed in Ref.~\cite{Durieux:2012gj}
require performing asymmetry measurements of same-sign dilepton pairs produced in
association with a top-quark jet. However, there is a process that could
give rise to a clean, unambiguous experimental signature:
baryon-antibaryon oscillations of hadrons that contain a valence quark from
each generation.
The only such baryon observed to date that decays weakly is
the \Xibz ($bsu$).
The interest of searching for \Xibz oscillations was noted
in Refs.~\cite{Kuzmin:1996sy,Bjorken:1991sg}, with an oscillation period potentially
as short as $\mathcal{O}(0.1\ps)$ suggested.
More recently, heavy baryon oscillations have been proposed as a
possible mechanism for baryogenesis~\cite{McKeen:2015cuz,Aitken:2017wie}.

\begin{figure}
  \begin{center}
    \includegraphics[width = \bnvsubplotwidth]{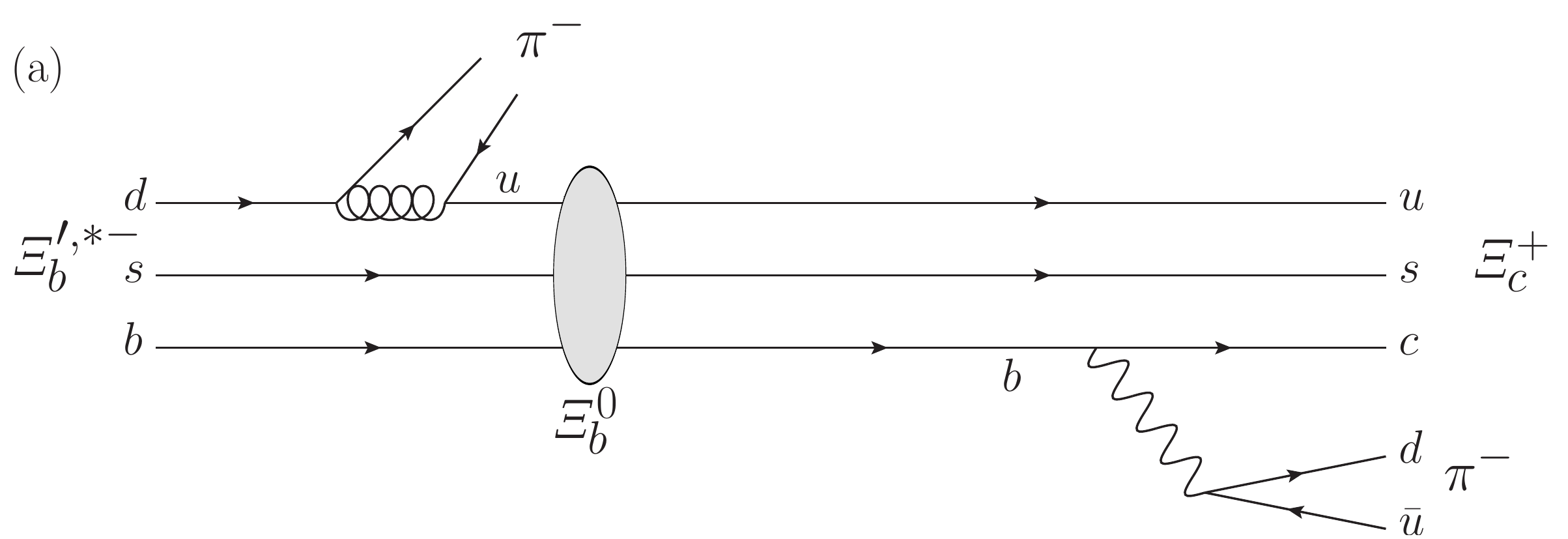}
    ~
    \includegraphics[width = \bnvsubplotwidth]{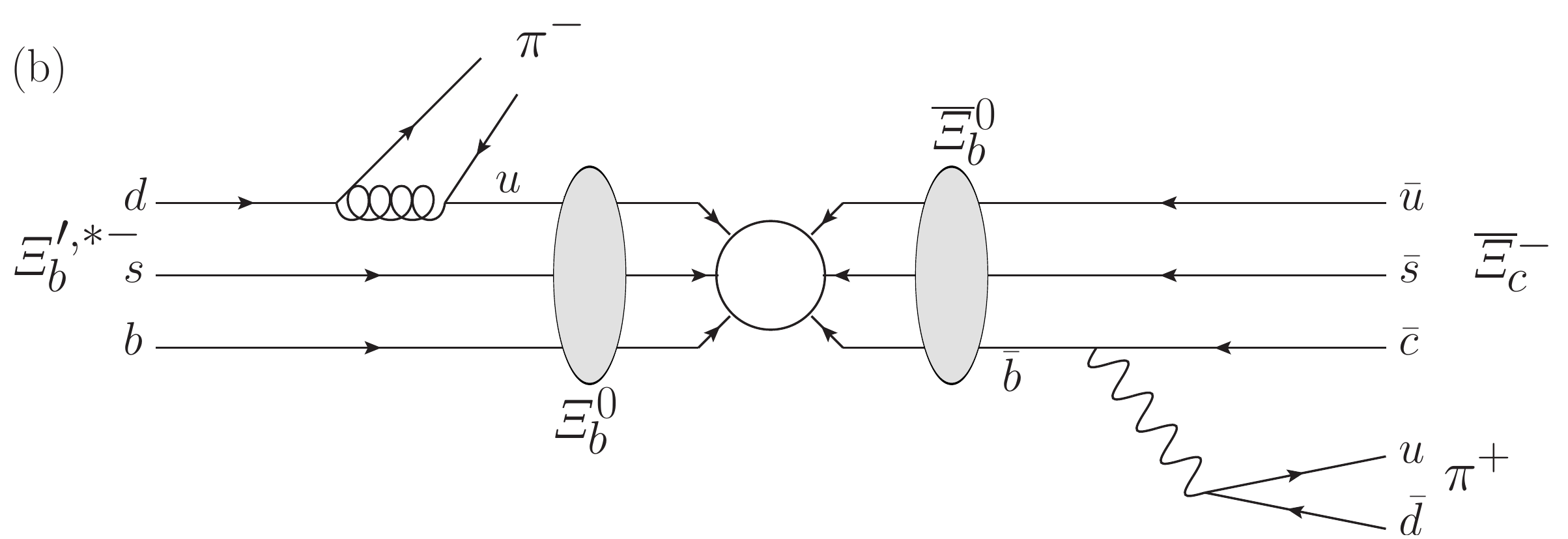}
  \end{center}
  \caption{
    (a) A non-BNV quark diagram for a $\XibPrimeOrStarMinus \to \Xibz \pim$
    strong decay followed by a $\Xibz \to \Xicp \pim$ weak decay.
    (b) The corresponding BNV diagram with a \Xibz to \Xibbarz oscillation followed by
    a decay to the final state $\Xicbarm \pip$.
  }
  \label{fig:feynman}
\end{figure}

The signature for a BNV process
is that a \Xibz
baryon is produced
and decays weakly as an antibaryon to a final state such as $\Xicbarm \pi^+$
(or, vice versa, that an antibaryon is produced and decays as a baryon).
The strong decays\footnote{
  The inclusion of charge-conjugate processes is implied throughout.
}
$\XibPrimeMinus \to \Xibz \pim$
and
$\XibStarMinus \to \Xibz \pim$
(denoted $\XibPrimeOrStarMinus \to \Xibz \pim$),
where \XibPrimeMinus and \XibStarMinus are the narrow resonances
$\XibxPrime(5935)^-$ and $\XibxStar(5955)^-$
recently observed by the LHCb collaboration~\cite{LHCb-PAPER-2014-061},
allow the baryon number at the time of production to be determined from
the charge of the pion.
Figure~\ref{fig:feynman} shows
quark-level
diagrams of example
(a)~non-BNV and (b)~BNV processes.

For baryon states propagating in free space, the
formalism for oscillations is similar to that of
neutral mesons~\cite{PDG2017,Harrison:1998yr,Bevan:2014iga},
which has been studied extensively in the context
of \Kz, \Dz, \Bz, and \Bs mixing~\cite{HFAG}.
However, a difference arises in the presence of
a magnetic field, $\vec{B}$, due to the 
nonzero magnetic moment, $\mu$,
possessed by the baryons,
resulting in a
splitting of the baryon and antibaryon energy levels
of $\Delta E = 2 \vec{\mu} \cdot \vec{B}$.
This splitting leads to a damping of the oscillations over time.
For the case
of neutron oscillations,
even a modest ambient magnetic
field would greatly suppress the oscillation probability
on the timescale of the neutron lifetime~\cite{Phillips:2014fgb}.
The criterion for the effect of the magnetic field to be negligible
is $|\Delta E| t / 2 \ll 1$,
where $t$ is the time of propagation of the baryon.
Taking the \Xibz magnetic moment to be comparable
to the nuclear magneton~\cite{PDG2017},
the energy splitting associated with the
magnetic field in the interaction region of the LHCb
detector,
which is $\lesssim 10$\,mT,
may be computed. For a typical time of propagation
equal to the known \Xibz lifetime~\cite{LHCb-PAPER-2014-021}
of $1.477 \pm 0.032$\ps,
$|\Delta E| t / 2 \lesssim 10^{-4}$.
This effect can therefore
be neglected and, in the limit of small mixing,
the ratio of the rate of oscillated decays, $P_{X \to \overline{X}}(t)$,
to the rate of nonoscillated decays, $P_{X \to X}(t)$,
varies over time as
\begin{equation}
  R(t) \equiv
  \frac{
  P_{X \to \overline{X}}(t)
  }{
  P_{X \to X}(t)
  } = \tan^2(t/\tau_{\mathrm{mix}}) 
  \simeq \frac{ t^2 }{ \tau_{\mathrm{mix}}^2 }
  \equiv (\omega t)^2 
  ,
  \label{eq:formalism:WSratio}
\end{equation}
where $2\pi\tau_{\mathrm{mix}}$ is the oscillation period, and
$\omega = 1/\tau_{\mathrm{mix}}$
gives the corresponding angular frequency
and is zero in the absence of oscillations.
This angular frequency is related to the mass difference $\Delta M$ and the
width difference $\Delta \Gamma$ between the eigenstates of the Hamiltonian
by
\mbox{$\omega^2 = (\Delta M /2)^2 + (\Delta \Gamma/4)^2$,}
and in the limit that BNV in the decay itself is negligible,
$\omega = \Delta M / 2$. 

This Letter presents a search for baryon-number-violating
\Xibz oscillations, performed with a sample of $\proton \proton$
collision data recorded by the
LHCb experiment, corresponding to an integrated luminosity of 3\invfb
collected at centre-of-mass energies $\sqrt{s}=7$\tev and
8\tev. This is the first such search for oscillations in heavy baryons.
The \lhcb detector~\cite{Alves:2008zz,LHCb-DP-2014-002} is a single-arm forward
spectrometer covering the \mbox{pseudorapidity} range $2<\eta <5$,
designed for the study of particles containing \bquark or \cquark
quarks.
The detector elements that are particularly
relevant to this analysis are: a silicon-strip vertex detector surrounding the $pp$ interaction
region that allows \cquark\ and \bquark\ hadrons to be identified from their characteristically long
flight distance; a tracking system that provides a measurement of momentum, $p$, of charged
particles; and two ring-imaging Cherenkov detectors that are able to discriminate between
different species of charged hadrons.
Samples of simulated events are used to study the
detector response and its effect on the measurement.
In the simulation, $pp$ collisions are generated using
\pythia~\cite{Sjostrand:2007gs,*Sjostrand:2006za} 
with a specific \lhcb
configuration~\cite{LHCb-PROC-2010-056}.  Decays of hadronic particles
are described by \evtgen~\cite{Lange:2001uf}, in which final-state
radiation is generated using \photos~\cite{Golonka:2005pn}. The
interaction of the generated particles with the detector, and its response,
are implemented using the \geant
toolkit~\cite{Allison:2006ve, *Agostinelli:2002hh} as described in
Ref.~\cite{LHCb-PROC-2011-006}.

Two classes of \XibPrimeOrStarMinus candidates are defined.
Baryon-number-conserving decays,
in which a strong decay $\XibPrimeOrStarMinus \to \Xibz \pim$
is followed by weak decays $\Xibz \to \Xicp \pim$ and $\Xicp \to \proton \Km \pip$,
are referred to as opposite-sign~(OS) candidates, since the \pim emitted
in the strong decay and the \proton have charges of opposite sign.
Conversely, in same-sign~(SS) candidates
the first decay $\XibPrimeOrStarMinus \to \Xibz \pim$
is followed by weak decays to a final state of different baryon number,
$\Xibbarz \to \Xicbarm \pip$ and $\Xicbarm \to \antiproton \Kp \pim$.

\begin{figure}
  \begin{center}
    \includegraphics[width = \bnvplotwidth]{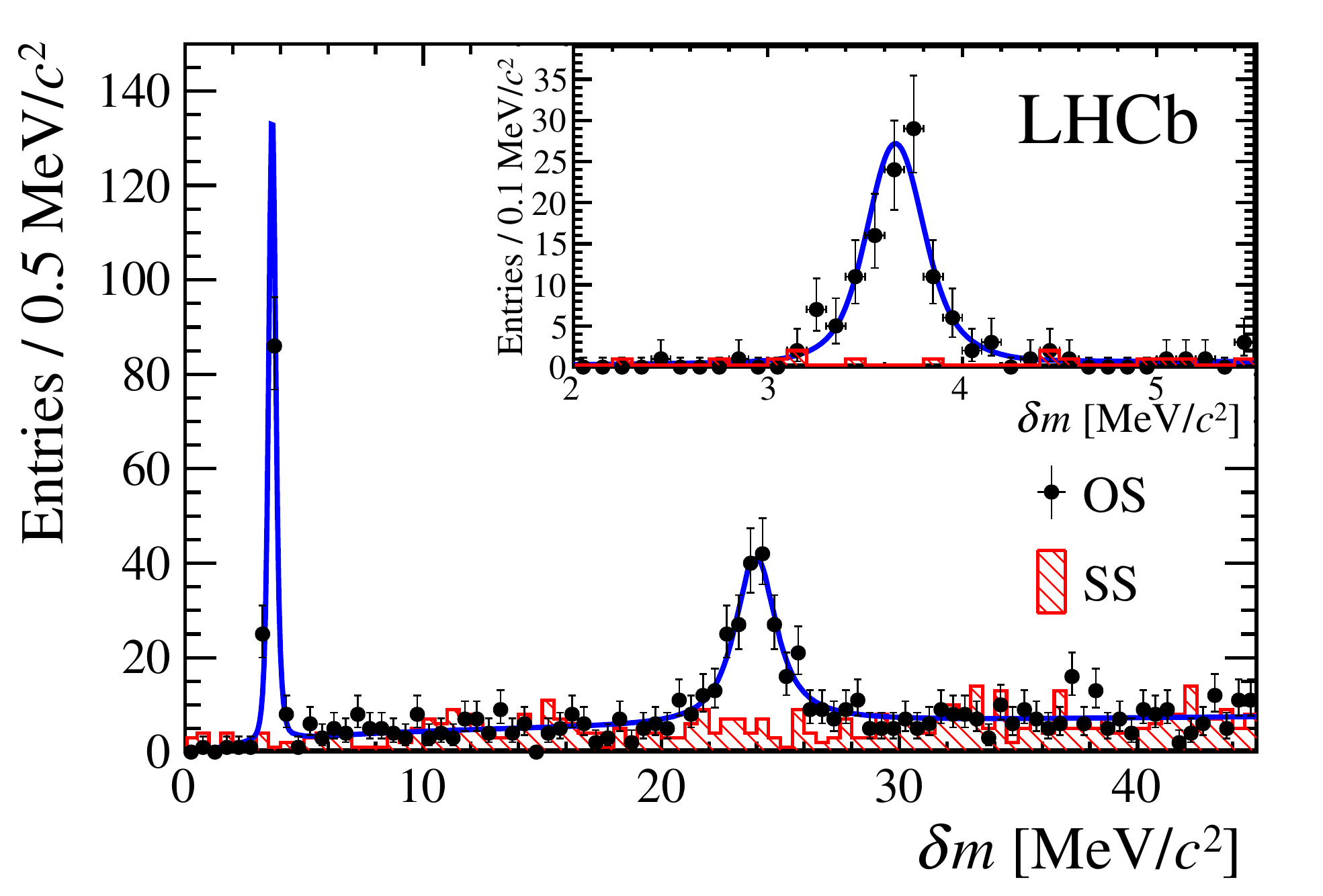}
  \end{center}
  \caption{
    Spectra of the mass difference
    $\delta m \equiv m(\Xibz \pi) - m(\Xibz) - m_\pi$
    in data after the full selection,
    for the opposite-sign (OS) sample (black points with error bars)
    and same-sign (SS) decays (red, hatched histogram).
    The blue curve is a fit to the OS data.
    The \XibPrimeMinus and \XibStarMinus peaks are at
    $\delta m \approx 3.7$ and $24$\mevcc, and the $\delta m$
    resolution at these points is approximately $0.2$ and $0.5$\mevcc, respectively;
    the \XibStarMinus also has a non-negligible natural width of $\Gamma \approx 1.7$\mev~\cite{LHCb-PAPER-2014-021}.
    Inset: detail of the region $2.0 < \delta m < 5.5$\mevcc.
  }
  \label{fig:mass}
\end{figure}

The reconstruction and selection procedures
are the same as those
described in Ref.~\cite{LHCb-PAPER-2014-061},
except for
one additional requirement on the track quality
of the pion produced in the $\XibPrimeOrStarMinus$ decay.
This requirement rejects a source of peaking background
that can arise when a genuine $\XibPrimeOrStarMinus \to \Xibz \pim$
decay occurs but the $\pim$ track is misreconstructed
such that its charge is incorrect and
the candidate migrates from the OS to the SS class~\cite{LHCb-PAPER-2015-057}.
In studies of simulated events, the fitted SS yield of this contribution is found to be
smaller than that of correctly reconstructed OS signal by a factor of
$(1.3 \pm 0.3) \times 10^{-3}$, where the uncertainty is statistical.
Applying the additional track quality requirement reduces the SS contribution in
simulation by an order of magnitude, such that it becomes smaller
than the OS yield by a factor of
$(1.6 \pm 2.0) \times 10^{-4}$,
corresponding to an expected SS peaking background yield of less than 0.1,
which is negligible.
The track quality requirement
also reduces the OS signal yield in data by approximately 10\%
and the combinatorial background by approximately 20\% compared to Ref.~\cite{LHCb-PAPER-2014-061}.
Figure~\ref{fig:mass}
shows the spectra of the
mass difference, $\delta m$, for the selected OS and SS candidates,
defining $\delta m \equiv m(\Xibz \pi) - m(\Xibz) - m_\pi$,
where $m_\pi$ is the known $\pi^{\pm}$ mass~\cite{PDG2017},
and $m(\Xibz \pi)$ and $m(\Xibz)$ are the reconstructed
invariant masses of the $\Xibz \pi$ and \Xibz candidates.
The figure also shows an unbinned extended maximum likelihood
fit to the OS candidates,
performed following the same procedure as described
below and in Ref.~\cite{LHCb-PAPER-2014-061}, as a blue curve.

\begin{figure}
  \begin{center}
    \includegraphics[width = \bnvplotwidth]{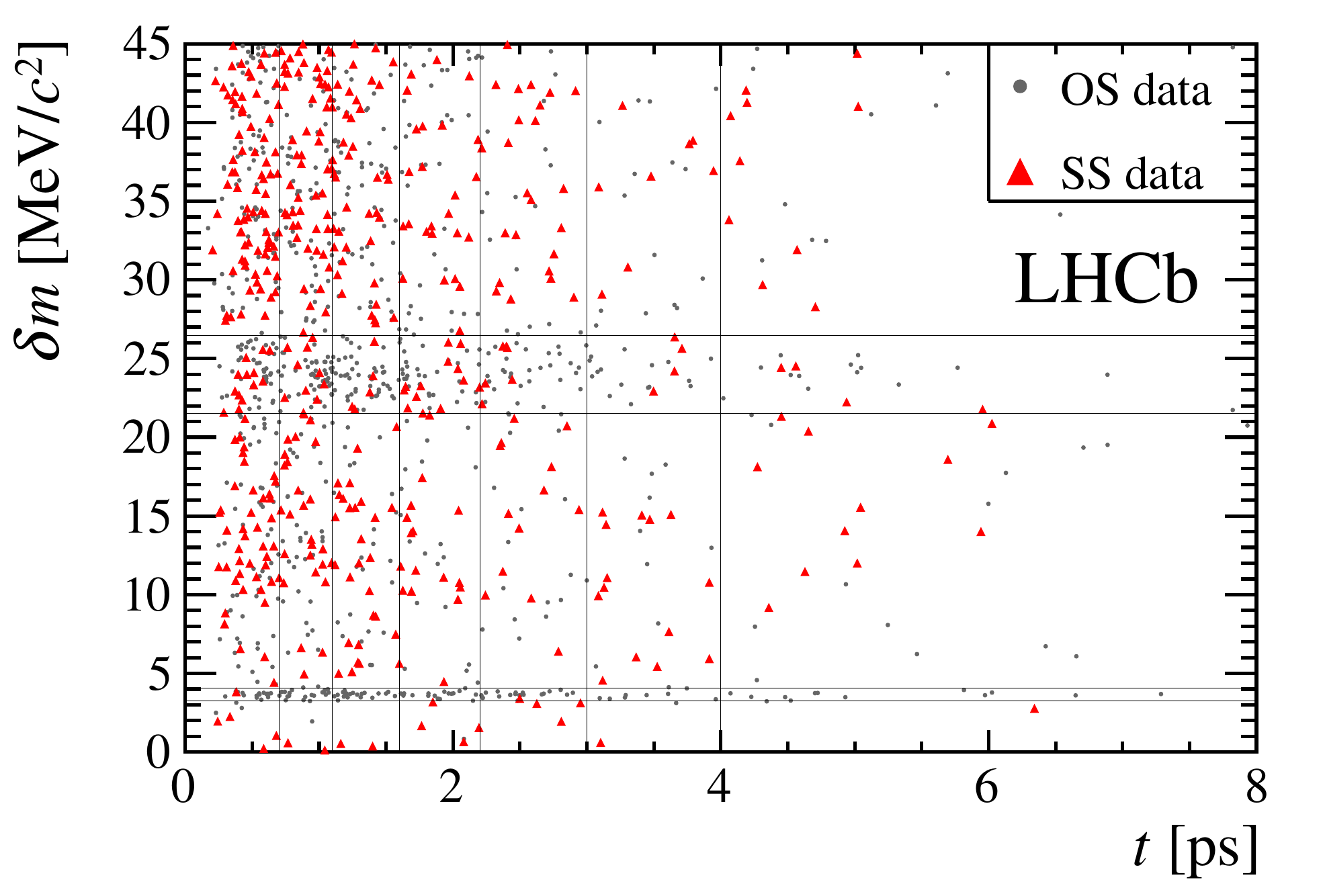}
  \end{center}
  \caption{
    Distribution of $\delta m$ vs decay time.
    The opposite-sign (OS) data are shown as grey points, and the
    same-sign (SS) data as larger red triangles.
    The vertical lines indicate the decay time bins.
    The horizontal lines are intended to guide the eye and
    indicate the \XibPrimeMinus and \XibStarMinus regions.
    Due to selection requirements, few candidates are present at
      short decay times.
  }
  \label{fig:scatter}
\end{figure}

The data are divided into seven bins of decay time 
(illustrated in Fig.~\ref{fig:scatter})
that have approximately equal OS signal yields
and cover the range $0 < t < 8$\ps, corresponding to approximately
$5.4$ times the mean \Xibz lifetime.
The OS resonance yields in the $i^{\mathrm{th}}$ bin are determined from
a fit to the $\delta m$ distribution of the OS data in that bin, with the resonance masses and
the \XibStarMinus width fixed to values obtained in a fit to the
whole OS data sample.
In each bin of decay time,
the shape and normalisation of the
SS combinatorial background are obtained from a fit to the $\delta m$
sideband regions of the SS data in that bin
(the sidebands being 0--2, 6--15, and 32--45\mevcc).
For a given value of the
angular frequency $\omega$ of the oscillations,
the expected ratio of SS to OS decays in the bin may be computed.
In combination with the OS yield
and the shape and normalisation of the SS background obtained as described above,
this fully determines the probability density function for the SS data in bin $i$,
and the corresponding likelihood
$L_i(\omega)$
is evaluated.
The overall likelihood is obtained by combining all bins as
$L(\omega) = \prod_i L_i(\omega)$. 

A test statistic $\Delta$ is  defined
based on the likelihood ratio approach as
$ { \Delta \equiv 2 \ln L(\hat{\omega}) - 2 \ln L(0) } $,
where $\hat{\omega}$ is the best-fit value of $\omega$
and is estimated from a likelihood scan.
Only the physical domain
$\omega \geq 0$ 
is considered, and consequently
$\hat{\omega}$
is expected to be zero approximately half of the time
under the null hypothesis.
The best-fit value
for the data
is found to be
$\hat{\omega}=0$
and the test statistic is therefore $\Delta = 0$.

Since no evidence of BNV oscillations is found,
an upper limit at the 95\% confidence level is placed on $\omega$
following the $\mathrm{CL}_s$
method~\cite{Read:2002hq,Cowan:2010js}.
Ensembles of parameterised simulations, referred to as pseudoexperiments, are generated
for a range of different oscillation
angular frequencies $\omega$.
The pseudoexperiments include
variation of efficiency with decay time,
decay time and
mass resolution,
combinatorial background,
and misclassification of OS candidates as SS via the misreconstruction described earlier.
To incorporate the associated systematic uncertainties, the
input parameters used to define the distributions
(the masses and yields of the resonances,
the natural width of the \XibStarMinus,
the background yield and shape parameters, and the signal misclassification rate)
are varied randomly within
their uncertainties for each pseudoexperiment.
Each pseudoexperiment is analysed in the same way as
data, and its test statistic $\Delta$ computed.
Coverage tests with pseudoexperiments indicate that the procedure
overcovers for small values of
$\omega$,
with 100\% coverage at
$\omega=0$,
and that the coverage converges
asymptotically to 95\% as the true value of
$\omega$
increases.

\begin{figure}
  \begin{center}
    \includegraphics[width = \bnvplotwidth]{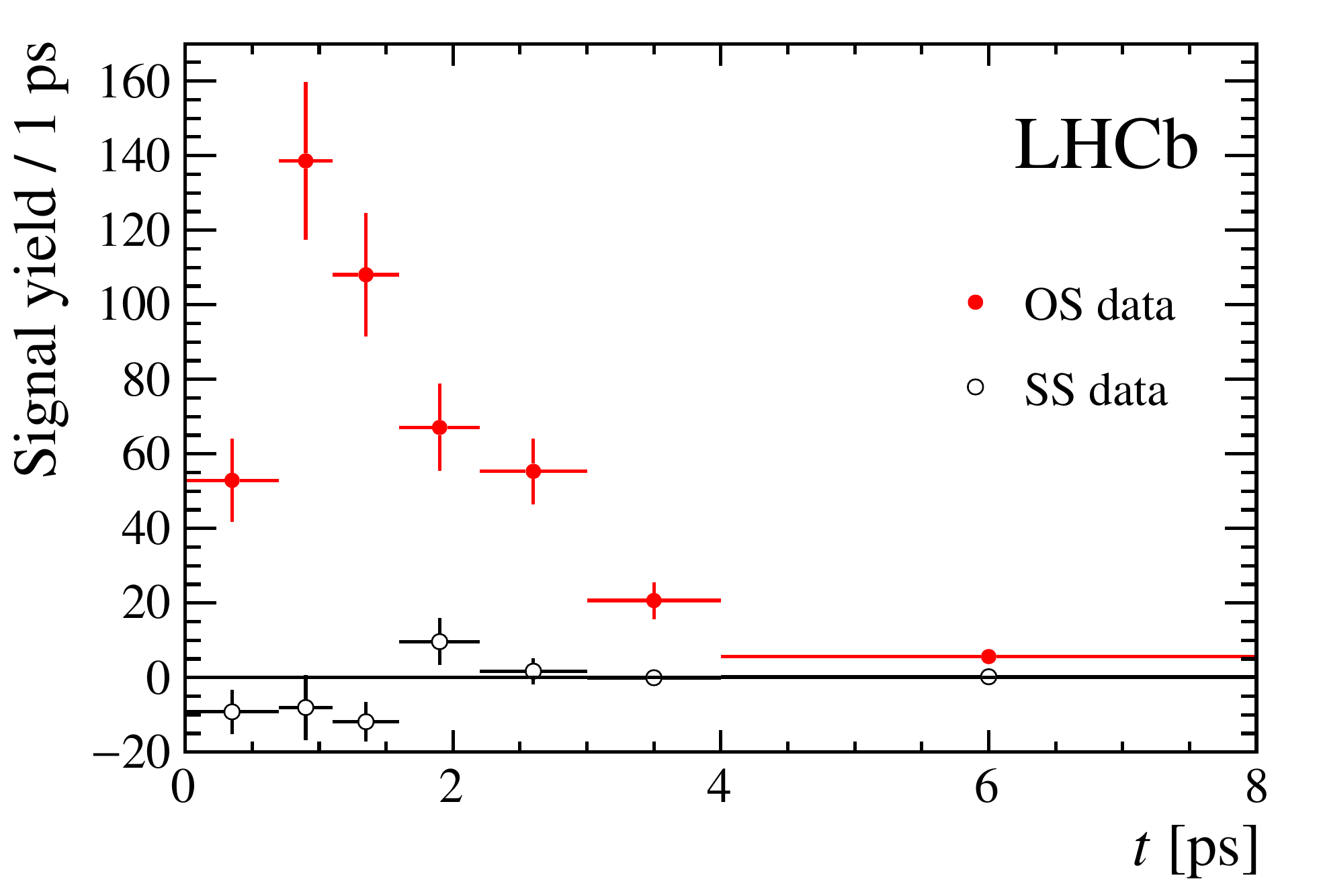}
    \includegraphics[width = \bnvplotwidth]{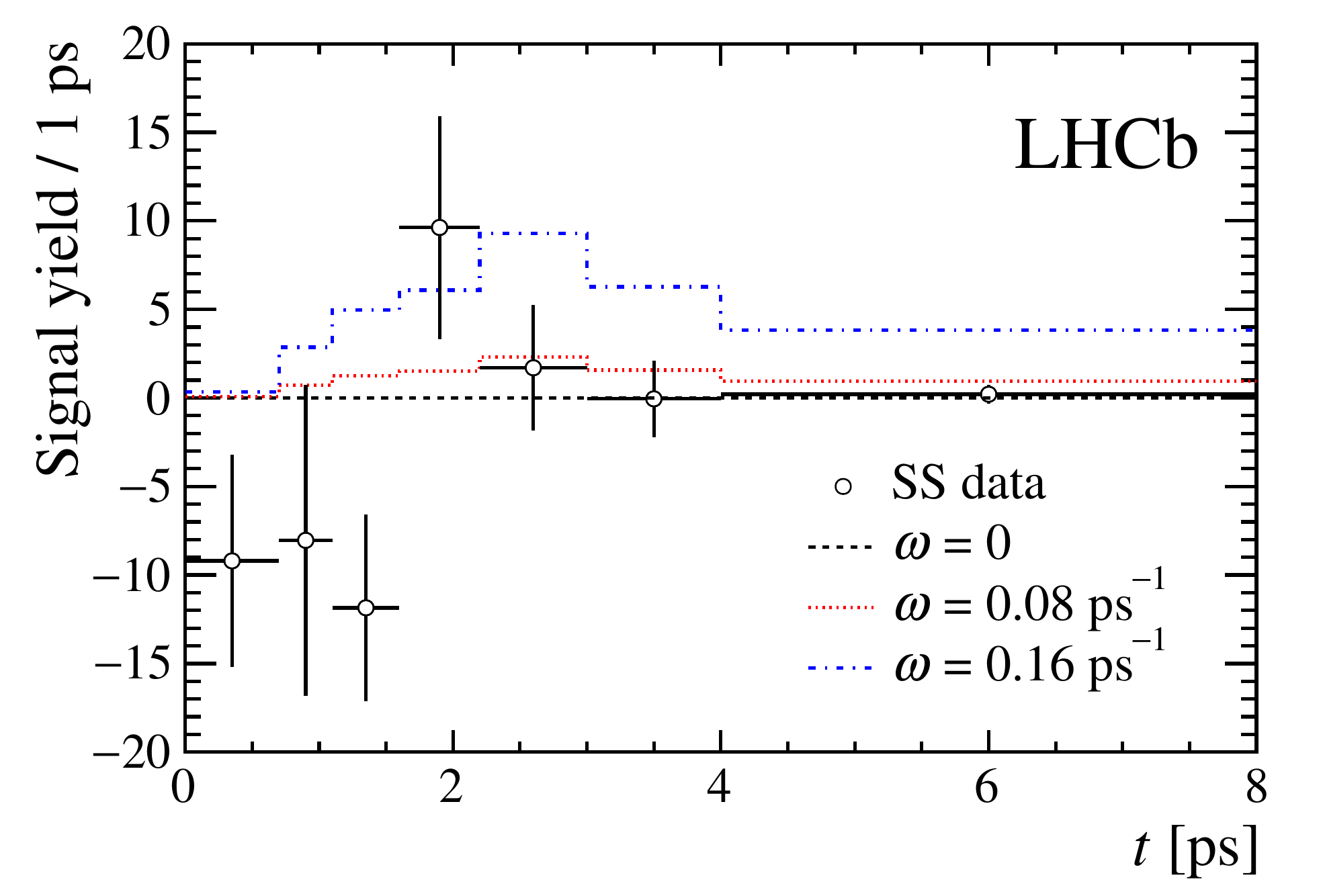}
  \end{center}
  \caption{
    For illustration, plots of the decay time distribution of signal candidates.
    In these plots,
    the two resonance regions are combined, and
    background is subtracted with a simplified statistical procedure that,
    unlike the likelihood described in the text, allows negative SS yields.
    (upper) The OS and SS data.
    (lower) Detail of the SS data, along with the expected SS yields
    under various hypotheses for the angular frequency $\omega$ of the oscillations.
  }
  \label{fig:decayTime}
\end{figure}

An upper limit of
$\omega < 0.08 \, \ps^{-1}$
at the 95\% confidence level is obtained,
which corresponds to
$\tau_{\mathrm{mix}} > 13$\ps.
This result can also be expressed in terms of the time-integrated
mixing rate $\chi$, defined as the fraction of particles produced
as \Xibz that decay as \Xibbarz, or vice versa.
Under the assumption of quadratic time dependence for $R(t)$,
$\chi = 2 \, \omega^2 \, \tau^2 < 2.7\%$ 
at the 95\% confidence level,
where $\tau$ is the known \Xibz lifetime~\cite{LHCb-PAPER-2014-021}.
For the purposes of illustration,
the evolution of the expected SS yield with decay time
for $\omega = 0.08 \, \ps^{-1}$
and $\omega = 0.16 \, \ps^{-1}$
is shown in
Fig.~\ref{fig:decayTime},
and compared to the SS yield in data as obtained with a simplified
statistical procedure.

In summary, a search is performed for baryon-antibaryon oscillations in the
\Xibz system. This is the first such search in the heavy-flavour sector, and is
of particular interest since \Xibz baryons
may couple directly to flavour-diagonal six-fermion operators
that violate baryon number~\cite{Durieux:2012gj}.
No evidence of baryon-number-violating oscillations is found.
In the limit of a small oscillation rate, the ratio of
same-sign to opposite-sign decays is expected to
increase quadratically with decay time.
A limit on the oscillation
angular frequency $\omega < 0.08 \, \ps^{-1}$ 
at the 95\% confidence level is obtained, equivalent to
$\tau_{\mathrm{mix}} > 13$\ps.
This rules out oscillations with a period comparable to the \Xibz
lifetime, as proposed in Ref.~\cite{Kuzmin:1996sy}.

%% file: acknowledgements.tex
%
%
\section*{Acknowledgements}

\noindent We express our gratitude to our colleagues in the CERN
accelerator departments for the excellent performance of the LHC. We
thank the technical and administrative staff at the LHCb
institutes. We acknowledge support from CERN and from the national
agencies: CAPES, CNPq, FAPERJ and FINEP (Brazil); MOST and NSFC
(China); CNRS/IN2P3 (France); BMBF, DFG and MPG (Germany); INFN
(Italy); NWO (The Netherlands); MNiSW and NCN (Poland); MEN/IFA
(Romania); MinES and FASO (Russia); MinECo (Spain); SNSF and SER
(Switzerland); NASU (Ukraine); STFC (United Kingdom); NSF (USA).  We
acknowledge the computing resources that are provided by CERN, IN2P3
(France), KIT and DESY (Germany), INFN (Italy), SURF (The
Netherlands), PIC (Spain), GridPP (United Kingdom), RRCKI and Yandex
LLC (Russia), CSCS (Switzerland), IFIN-HH (Romania), CBPF (Brazil),
PL-GRID (Poland) and OSC (USA). We are indebted to the communities
behind the multiple open-source software packages on which we depend.
Individual groups or members have received support from AvH Foundation
(Germany), EPLANET, Marie Sk\l{}odowska-Curie Actions and ERC
(European Union), ANR, Labex P2IO, ENIGMASS and OCEVU, and R\'{e}gion
Auvergne-Rh\^{o}ne-Alpes (France), RFBR and Yandex LLC (Russia), GVA,
XuntaGal and GENCAT (Spain), Herchel Smith Fund, the Royal Society,
the English-Speaking Union and the Leverhulme Trust (United Kingdom).

%% file: LHCb_Authorship_flat_14-Jun-2017.tex
\centerline{\large\bf LHCb collaboration}
\begin{flushleft}
\small
R.~Aaij$^{40}$,
B.~Adeva$^{39}$,
M.~Adinolfi$^{48}$,
Z.~Ajaltouni$^{5}$,
S.~Akar$^{59}$,
J.~Albrecht$^{10}$,
F.~Alessio$^{40}$,
M.~Alexander$^{53}$,
A.~Alfonso~Albero$^{38}$,
S.~Ali$^{43}$
% ...

{\footnotesize \it
% ...
$ ^{5}$Clermont Universit{\'e}, Universit{\'e} Blaise Pascal, CNRS/IN2P3, LPC, Clermont-Ferrand, France\\
% ...
$ ^{10}$Fakult{\"a}t Physik, Technische Universit{\"a}t Dortmund, Dortmund, Germany\\
% ...
$ ^{38}$ICCUB, Universitat de Barcelona, Barcelona, Spain\\
$ ^{39}$Universidad de Santiago de Compostela, Santiago de Compostela, Spain\\
$ ^{40}$European Organization for Nuclear Research (CERN), Geneva, Switzerland\\
% ...
$ ^{43}$Nikhef National Institute for Subatomic Physics, Amsterdam, The Netherlands\\
% ...
$ ^{48}$H.H. Wills Physics Laboratory, University of Bristol, Bristol, United Kingdom\\
% ...
$ ^{53}$School of Physics and Astronomy, University of Glasgow, Glasgow, United Kingdom\\
% ...
$ ^{59}$University of Cincinnati, Cincinnati, OH, United States\\
% ...
}
\end{flushleft}